\documentclass[letterpaper, 10 pt, conference]{ieeeconf}  

\IEEEoverridecommandlockouts                              

\usepackage{cite}
\usepackage{amsmath,amssymb,amsfonts,bm}
\usepackage{algorithmic}
\usepackage{graphicx}
\usepackage{textcomp}
\usepackage{epsfig}
\usepackage{times}
\usepackage{longtable}
\usepackage{supertabular,booktabs}
\usepackage{ltablex}
\usepackage{array}
\usepackage{setspace}
\usepackage{xcolor,soul}
\usepackage{balance}
\usepackage{etoolbox}
\usepackage[labelsep=period,font={small}]{caption}

\newcolumntype{C}[1]{>{\centering\arraybackslash} m{#1}}

\newcommand{\qed}{\hfill $\square$}

\renewcommand{\thetable}{\arabic{table}}

\title{\LARGE \bf
CBF Based Quadratic Program for Trajectory Tracking of Underatuated Marine Vessels*}

\author{Ji-Hong Li 
\thanks{*This work was supported in part by Korea Institute of Marine Science \& Technology Promotion (KIMST) funded by the Ministry of Oceans and Fisheries (RS-2023-00256122), also in part by KIMST funded by the Ministry of Oceans and Fisheries (RS-2024-00432366), all in the Republic of Korea.}
\thanks{J. H. Li is with the Autonomous Systems R\&D Division, Korea Institute of Robotics and Technology Convergence,
        Jigok-Ro 39, Nam-Gu, Pohang 37666, Republic of Korea {\tt\small \{jhli5\}@kiro.re.kr}}%
}

\begin{document}

\maketitle
\thispagestyle{empty}
\pagestyle{empty}

\begin{abstract}
By introducing two polar coordinates transformations, the marine vessel's original two-input-three-output second-order tracking model can be reduced to a two-input-two-output feedback form. However, the resulting system does not confirm to the strict-feedback structure, leading to potential singularity when designing the stabilizing function for the virtual input in the recursive controller design. Moreover, the polar coordinate transformation itself inherently introduces singularities. To address these singularity issues, this paper employs a control barrier function (CBF) based approach and formulates the trajectory tracking problem as a quadratic program (QP) solved via a QP optimizer. Numerical simulations are carried out to demonstrate the effectiveness of the proposed method.
\end{abstract}

\section{Introduction}
Over the past decades, the trajectory tracking of underactuated marine vessels has become one of the most active research topic in the nonlinear control community. Related works can be roughly classified into two groups: trajectory tracking and path following \cite{b1}--\hspace{1sp}\cite{b5}. Depending on how the vessel's reference yaw angle is defined, the former trajectory tracking can be further subdivided into two categories \cite{b6,b7}. If the reference yaw angle is predefined, the problem corresponds to traditional trajectory tracking; otherwise, it corresponds to trajectory following. From a control design perspective, the former would be more challenging.

This paper focuses on the traditional trajectory tracking problem. In this context, most existing methods \cite{b3,b8}--\hspace{1sp}\cite{b11} impose various constraints on the given reference trajectory and, in some cases, on the controller design parameters. In addition to applying simplified vessel models, some approaches require restrictive conditions, such as enforcing that the yaw angular rate satisfies the persistency of excitation (PE) condition \cite{b3,b8}. In \cite{b10}, a nonlinear sliding mode control law was proposed with the vessel's general dynamics. However, the presented method is only applicable in the case where the reference trajectory is non-straight-line. To overcome these restrictions, in \cite{b12}, two polar coordinate transformations were introduced to convert the vessel's original two-input-three-output second-order system into a two-input-two-output feedback form, enabling the tracking problem to be addressed using the general backstepping method \cite{b13}. Unfortunately, the resulting transformed system does not confirm to the strict-feedback form, which can lead to singularities when designing the stabilizing function in the recursive controller design. To avoid this potential singularity, the asymptotic modification of orientation (AMO) concept was introduced in \cite{b12}. This concept, along with similar approaches such as the exponential modification of orientation (EMO) in \cite{b6}, has been adopted in several subsequent works \cite{b7,b14,b15}.

In this paper, rather than applying AMO or EMO concepts as in \cite{b6,b7,b12,b14,b15}, we address the tracking problem and aim to avoid singularities by employing a control barrier function (CBF) method. Recently, CBFs have been widely used not only for system safety but also for singularity avoidance in both robotics and control applications \cite{b16}--\hspace{1sp}\cite{b20}. Through proper CBF design, various practical singularities can be avoided, which serves as a primary motivation for this study. On the other hand, for the completeness of the sideslip angle definition, the vessel's surge speed is always assumed to be strictly positive during tracking in \cite{b6,b7,b12,b14,b15}. This restrictive assumption can be removed in this paper by simply including it as a constraint in the QP solver, which constitutes another main contribution of this paper.

The remainder of this paper is organized as follows. Section II presents the vessel model, two polar coordinate transformations, and other relevant preliminaries. In particular, various singularity issues arising from the system transformations are discussed in detail. The main results are provided in Section III, while Section IV demonstrates the effectiveness of the proposed method through numerical simulations. Finally, Section V concludes the paper with a brief summary.

\section{Problem Statement}
\subsection{Vessel Model}
This paper considers the marine vessels with the following kinematics and dynamics \cite{b5,b7}
\begin{align}
\begin{bmatrix}\dot{x}\\ \dot{y}\\ \dot{\psi}\end{bmatrix}&=\begin{bmatrix}\cos\psi&-\sin\psi&0\\ \sin\psi&\cos\psi&0\\0&0&1\end{bmatrix}\begin{bmatrix}u\\v\\r\end{bmatrix}, \label{eq1} \\
\begin{bmatrix}\dot{u}\\ \dot{v}\\ \dot{r}\end{bmatrix}&=\begin{bmatrix}f_u\\f_v\\f_r\end{bmatrix}+\begin{bmatrix}b_u&0\\0&\epsilon_r\\0&b_r\end{bmatrix} \begin{bmatrix}\tau_u\\ \tau_r\end{bmatrix}, \label{eq2}
\end{align}
where $(x,y)$ denote the horizontal position and $\psi$ is the yaw angle, all defined in the navigation frame; $u$ and $v$ represent the surge and sway velocities in the body-fixed frame, respectively, and $r$ is the yaw rate in the inertial frame; $f_u,f_v,f_r\in C^2$ describe the modeled nonlinear dynamics of the vessel, incorporating hydrodynamic damping, inertia (including added mass), Coriolis and centripetal, and gravitational effects in the surge, sway, and yaw directions; the surge force $\tau_u$ and the yaw moment $\tau_r$ are the only available control inputs, where $b_u$ and $b_r$ are nonzero constants, and $\epsilon_r$ represents the lift effect induced by the yaw moment \cite{b22}.

\emph{Remark 1}. In (\ref{eq2}), the system is minimum phase when $\epsilon_r=0$ and non-minimum phase otherwise \cite{b22,b23}. Both cases are considered in this paper. It is noteworthy that the minimum phase property does not affect the design of the proposed method.

\subsection{Two Polar Coordinates Transformations}
The first polar coordinate transformation applied in this paper is that \cite{b7,b12}
\begin{equation}
\begin{bmatrix}u_l\\ \psi_a\end{bmatrix}=\mathcal{F}_a(u,v):=\begin{bmatrix}\sqrt{u^2+v^2}\\ \arctan(v/u)\end{bmatrix}, \label{eq3}
\end{equation}
where $\psi_a$ is commonly known as the sideslip angle \cite{b20}.

With (\ref{eq3}), the vessel's dynamics (\ref{eq2}) can be reformulated as following
\begin{equation}
\begin{bmatrix}{\dot{u}}_l \\ \dot{r}_l\end{bmatrix}=\begin{bmatrix}f_{u_l} \\ f_{r_l}\end{bmatrix}+\begin{bmatrix}b_{u_l} &\epsilon_{ra} \\0 &b_r\end{bmatrix}\begin{bmatrix}\tau_u\\ \tau_r\end{bmatrix}, \label{eq4}
\end{equation}
where $r_l=r+\dot{\psi}_a$, $f_{u_l}=\cos\psi_a f_u+\sin\psi_a f_v$, $f_{r_l}=f_r+\ddot{\psi}_a$, $b_{u_l}=\cos\psi_a b_u$, $\epsilon_{ra}=\sin\psi_a\epsilon_r$.

Corresponding to $r_l$ defined in (\ref{eq4}), if we define $\dot{\psi}_l=r_l$ (which implies $\psi_l=\psi+\psi_a$), the vessel's kinematics can then be expressed as follows:
\begin{equation}
\left\{\begin{matrix}\dot{x}=u_l\cos\psi_l, \\ \dot{y}=u_l\sin\psi_l.\end{matrix}\right. \label{eq5}
\end{equation}

Given a reference trajectory $\bm{\eta}_{ld}(t)=(x_d(t),y_d(t),\psi_{ld}(t))$ with $\dot{x}_d=u_{ld}\cos\psi_{ld}$ and $\dot{y}_d=u_{ld}\sin\psi_{ld}$, the following second polar coordinate transformation \cite{b6,b11} will be employed in this paper:
\begin{equation}
\begin{bmatrix}p_e\\ \psi_b\end{bmatrix}=\mathcal{F}_b(x_e,y_e):=\begin{bmatrix}\sqrt{x_e^2+y_e^2}\\ \text{atan2}(y_e,x_e)\end{bmatrix}, \label{eq6}
\end{equation}
where $x_e=x_d-x$, $y_e=y_d-y$, and $\psi_b$ is defined as the azimuth angle from the vessel to the target point moving on the reference trajectory.

According to (\ref{eq6}), the time derivative of $p_e(t)$ becomes
\begin{align}
\dot{p}_e&=\frac{x_e}{p_e}(\dot{x}_d-\dot{x})+\frac{y_e}{p_e}(\dot{y}_d-\dot{y}) \nonumber \\
&=u_{ld}\cos(\psi_{ld}-\psi_b)-u_l \cos(\psi_l -\psi_b). \label{eq7}
\end{align}

Consequently, the vessel's tracking error kinematics can be derived as follows:
\begin{equation}
\begin{bmatrix}\dot{p}_e\\ \dot{\psi}_{le}\end{bmatrix}\!=\!\begin{bmatrix}u_{ld}\cos(\psi_{ld}\!-\!\psi_b)\\ \dot{\psi}_{ld} \end{bmatrix}\!-\!\begin{bmatrix}\cos(\psi_l\!-\!\psi_b) &0\\0 &1\end{bmatrix}\!\!\begin{bmatrix}u_l \\r_l\end{bmatrix}. \label{eq8}
\end{equation}
where $\psi_{le}=\psi_{ld}-\psi_l$.

Combining (\ref{eq8}) and (\ref{eq4}), finally we can get the following two-input-two-output second-order feedback form of the trajectory tracking error model:
\begin{align}
\begin{bmatrix}\dot{p}_e\\ \dot{\psi}_{le}\end{bmatrix}\!&=\!\begin{bmatrix}u_{ld}\cos(\psi_{ld}\!-\!\psi_b)\\ \dot{\psi}_{ld} \end{bmatrix}\!-\!\begin{bmatrix}\cos(\psi_l\!-\!\psi_b) &0\\0 &1\end{bmatrix}\!\!\begin{bmatrix}u_l \\r_l\end{bmatrix}, \label{eq9} \\
\begin{bmatrix}{\dot{u}}_l \\ \dot{r}_l\end{bmatrix}\!&=\!\begin{bmatrix}f_{u_l} \\ f_{r_l}\end{bmatrix}+\begin{bmatrix}b_{u_l} &\epsilon_{ra} \\0 &b_r\end{bmatrix}\begin{bmatrix}\tau_u\\ \tau_r\end{bmatrix}. \label{eq10}
\end{align}

\subsection{System Singularities}
Considering the second-order two-input-two-output feedback form of the tracking system (\ref{eq9}) and (\ref{eq10}), if it were in strict-feedback form, the tracking problem could be readily solved using the standard backstepping method \cite{b13}. However, in practice, this assumption does not hold, as the coefficients of both the virtual input $u_l$ and the actual control input $\tau_u$ may become zero. Under such conditions, certain singularities can arise during the recursive controller design. For clarity, these singular points (SPs) are listed as follows:
\begin{itemize}
\setlength{\itemindent}{20pt}
\item [SP-1.] $\cos(\psi_l-\psi_b)=0$.
\item [SP-2.] $b_{u_l}=0$.
\end{itemize}

Moreover, the polar coordinate system inherently introduces a singularity, since the polar angle is undefined at the origin. Due to the application of two polar coordinates transformations, it becomes necessary to consider the following additional singular points:
\begin{itemize}
\setlength{\itemindent}{20pt}
\item [SP-3.] $\psi_a$ is undefined when $u_l=0$.
\item [SP-4.] $\psi_b$ is undefined when $p_e=0$.
\end{itemize}

\emph{Remark 2}. To avoid the singularity of SP-1, for a given $(u_{ld},\psi_{ld})$, the authors introduced the concepts AMO \cite{b7,b12,b14,b15} and EMO \cite{b6}, denoted by $(u_{ld}^m,\psi_{ld}^m)$, which ensure asymptotic or exponential convergence of the position error $p_e(t)$. Here, the modified orientation $u_{ld}^m$ is directly taken as the stabilizing function for the virtual input $u_l$. In \cite{b14}, the azimuth angle $\psi_b$ was directly taken as the vessel's desired heading for $\psi_l$, which also allows SP-1 to be avoided.

\emph{Remark 3}. Since $b_{u_l}=b_u\cos\psi_a$ with $b_u\neq0$, $b_{u_l}=0$ is equivalent to $\cos\psi_a=0$, which corresponds to the case where $u=0$ and $v\neq0$. To avoid the singularities SP-2 and SP-3, in \cite{b6,b7,b12,b14,b15}, the vessel's surge velocity $u(t)$ was assumed to satisfy $u(t)>0$, $\forall t>0$.

\emph{Remark 4}. In the case of the fourth singularity, SP-4, it is noteworthy that it does not affect the design of the proposed controller in \cite{b6,b7,b12,b14,b15}. Nevertheless, in practical applications, when $p_e$ becomes very small, the computed $\psi_b$ is heavily influenced by measurement noise, which can significantly compromise control performance. To address this issue, \cite{b14} replaced $p_e\rightarrow 0$ with $p_e\rightarrow c_d$, where $c_d>0$ is a design parameter. This modification also explains why the tracking method in \cite{b14} is referred to as towing-type trajectory tracking.

\subsection{Problem Formulation}
Unlike the existing methods reported in \cite{b6,b7,b12,b14,b15}, this paper aims to introduce a CBF-based approach to effectively avoid the above singularities SP-1$\sim$4.

It is straightforward to formulate the constraint condition (CC) for avoiding the singularity SP-1 as follows:
\begin{itemize}
\setlength{\itemindent}{20pt}
\item [CC-1.] $|\cos[\psi_l(t)-\psi_b(t)]|>0$, $\forall t\geq0$.
\end{itemize}

As for avoiding SP-2 and SP-3, the necessary condition is that $u(t)\neq 0$, $\forall t\geq0$. Without loss of generality, this paper considers the following constraint:
\begin{itemize}
\setlength{\itemindent}{20pt}
\item [CC-2.] $u(t)>0$, $\forall t\geq0$.
\end{itemize}

\emph{Remark 5}. For most marine vessels, a single thruster at the stern generates the surge force, while one or more rudders control the yaw dynamics. In such configurations, when $u(t)=0$, the yaw moment also becomes zero. Consequently, in order to treat the only available control inputs, $\tau_u$ and $\tau_r$, as independent variables, the vessel must maintain a sufficiently large forward speed. Therefore, $u(t)>0$ is not only necessary to avoid SP-2 and SP-3, but also practically required for effective control implementation.

Finally, as a measure to avoid SP-4, the method proposed in \cite{b14}, namely replacing $p_e\rightarrow 0$ with $p_e\rightarrow c_d>0$, is directly adopted. In this case, the substitution $p_e\rightarrow c_d$ is incorporated into the Lyapunov function selection for designing the reference controller $\bm{\tau}^{ref}$. Consequently, the control objective in this paper is reduced to proposing a solver for the following convex QP:
\begin{align}
\underset{\dot{\bm{\nu}}_l,\bm{\tau}}{\text{min}}~~&||\bm{\tau}-\bm{\tau}^{r\!e\!f}||^2 \label{eq11} \\
\text{s.t.}~~&\text{Constraints CC-1 and CC-2}, \nonumber
\end{align}
where $\bm{\tau}=[\tau_u,\tau_r]^T$, and $\bm{\tau}^{r\!e\!f}$ denotes the reference control input, which will be designed later following the proposed method in \cite{b6,b7,b12,b14,b15}; $\bm{\nu}_l=[u_l,r_l]^T$, and $\bm{f}_l=[f_{u_l},f_{r_l}]^T$.

\section{Main Results}
This section presents the main results. However, before that, it is necessary to derive the required CBFs from CC-1 and CC-2 to avoid the system singularities SP-1$\sim$4.

\subsection{Control Barrier Functions for Singularities Avoidance}
\subsubsection{CBF for CC-1}
It is noteworthy that all yaw-related quantities, including $\psi_l-\psi_b$, are modulated within the domain $[-\pi,\pi)$. Regarding CC-1, two different cases are considered for the modulated term $\psi_l-\psi_b$. The first corresponds to $-\pi/2<\psi_l-\psi_b<\pi/2$, for which the following barrier function is proposed:
\begin{equation}
h(\bm{\eta}_l)=\cos(\psi_l-\psi_b)-\epsilon_{\psi}, \label{eq12}
\end{equation}
where $\bm{\eta}_l=(x,y,\psi_l)$, and $\epsilon_{\psi}>0$ is a constant design parameter characterizing a minimum ``distance'' from singularities \cite{b20}.

For this relative degree 2 of barrier function $h(\bm{\eta}_l)$, corresponding exponential CBF (ECBF) condition is that \cite{b17}--\hspace{1sp}\cite{b20},
\begin{equation}
\ddot{h}(\bm{\eta}_l,\bm{\nu}_l,\dot{\bm{\nu}}_l)\geq -\bm{K}_{\alpha}\begin{bmatrix}h(\bm{\eta}_l)\\ \dot{h}(\bm{\eta}_l,\bm{\nu}_l)\end{bmatrix}, \label{eq13}
\end{equation}
where $\dot{h}(\bm{\eta}_l)=-\sin(\psi_l-\psi_b)(\dot{\psi}_l-\dot{\psi}_b)$, $\ddot{h}(\bm{\eta}_l,\bm{\nu}_l,\dot{\bm{\nu}}_l)=-\cos(\psi_l-\psi_b)(\dot{\psi}_l-\dot{\psi}_b)^2 -\sin(\psi_l-\psi_b)(\ddot{\psi}_l-\ddot{\psi}_b)$, and $\bm{K}_{\alpha}=[\alpha_1, \alpha_2]$ satisfies the conditions of Theorem 2 in \cite{b17} and Theorem 8 in \cite{b19}, such that $\bm{F}-\bm{G}\bm{K}_a$ is Hurwitz, where
\begin{equation*}
\bm{F}=\begin{bmatrix}0&1\\0&0\end{bmatrix},~~~\bm{G}=\begin{bmatrix}0\\1\end{bmatrix}.
\end{equation*}

Combining (\ref{eq12}) and (\ref{eq13}), and omitting the laborious and detailed numerical expansions, we obtain the following inequality
\begin{equation}
\begin{bmatrix}c_1\!&\!c_2\end{bmatrix}\begin{bmatrix}\tau_u\\ \tau_r\end{bmatrix}\leq \sin(\psi_l-\psi_b)M-\alpha_1\epsilon_{\psi}, \label{eq14}
\end{equation}
where
\begin{align*}
c_1\!=&\dfrac{b_{u_l}\sin^2(\psi_l-\psi_b)}{p_e}, \\
c_2\!=&\sin(\psi_l-\psi_b)\left[\dfrac{\epsilon_{ra}\sin(\psi_l-\psi_b)}{p_e}+b_r\right], \\
M\!=&\left[\dot{u}_{ld}\sin(\psi_{ld}-\psi_b)+u_{ld}\cos(\psi_{ld}-\psi_b)(\dot{\psi}_{ld}-\dot{\psi}_b) \right. \\
&\left.-u_l\cos(\psi_l\!-\!\psi_b)(\dot{\psi}_l\!-\!\dot{\psi}_b)\!-\!\!f_{u_l}\!\sin(\psi_l\!-\!\psi_b)\!-\!\dot{\psi}_b\dot{p}_e\right]\!/\!p_e \\
&-\!f_{r_l}+\alpha_1\cos(\psi_l-\psi_b)-\alpha_2\sin(\psi_l-\psi_b)(\dot{\psi}_l-\dot{\psi}_b).
\end{align*}

Next, consider the second case where $\psi_l-\psi_b>-\pi/2$ or $\psi_l-\psi_b>-\pi/2$, for which the corresponding barrier function is proposed as follows:
\begin{equation}
h(\bm{\eta}_l)=-\cos(\psi_l-\psi_b)-\epsilon_{\psi}. \label{eq15}
\end{equation}

Through a similar expansion process as described above, the following inequality is obtained:
\begin{equation}
\begin{bmatrix}-c_1\!&\!-c_2\end{bmatrix}\begin{bmatrix}\tau_u\\ \tau_r\end{bmatrix}\leq -\sin(\psi_l-\psi_b)M-\alpha_1\epsilon_{\psi}, \label{eq16}
\end{equation}

\emph{Proposition 1}. ECBF condition for CC-1 can be formulated as follows:
\begin{equation}
\begin{array}{ll}
\text{if} & -\pi/2<\psi_l-\psi_b<\pi/2\\
&\begin{bmatrix}c_1\!&\!c_2\end{bmatrix}\bm{\tau}\leq \sin(\psi_l-\psi_b)M-\alpha_1\epsilon_{\psi}, \\
\text{elseif} & \psi_l-\psi_b>\pi/2~\text{or}~\psi_l-\psi_b<-\pi/2\\
&\begin{bmatrix}-c_1\!&\!-c_2\end{bmatrix}\bm{\tau}\leq\!-\sin(\psi_l\!-\!\psi_b)M\!-\!\alpha_1\epsilon_{\psi}.\end{array} \label{eq17}
\end{equation}

\emph{Proof}. Given the reference trajectory $\bm{\eta}_{ld}(t)$ and the vessel's motion information, it is always feasible to calculate $M$ as defined in (\ref{eq14}). Moreover, to ensure that $\bm{F}-\bm{G}\bm{K}_a$ is Hurwitz, it follows that $\alpha_1>0$. Therefore, for given $c_1$ and $c_2$, through proper design of $\bm{\tau}=[\tau_u,\tau_r]^T$ and $\epsilon_{\psi}$, it is feasible for us to satisfy the conditions in (\ref{eq17}), which in turn ensures that both (\ref{eq12}) and (\ref{eq15}) are ECBF. Therefore, the proposition holds. \qed

Since (\ref{eq17}) is linear in control law $\bm{\tau}$, it can be included in a QP like (\ref{eq11}). To match the constraint format used in \emph{quadprog($\cdot$)} function provided by MATLAB, (\ref{eq17}) can be rewritten in the following form
\begin{equation}
\begin{array}{ll}
\text{if} & -\pi/2<\psi_l-\psi_b<\pi/2\\
&\begin{bmatrix}c_1\!&\!c_2\end{bmatrix}\!\bm{X}\!\leq \sin(\psi_l-\psi_b)M-\alpha_1\epsilon_{\psi} \\
&~~~~~~~~~\!~~~~~~~-\begin{bmatrix}c_1&c_2\end{bmatrix}\bm{\tau}^{r\!e\!f}, \\
\text{elseif} & \psi_l-\psi_b>\pi/2~\text{or}~\psi_l-\psi_b<-\pi/2\\
&\begin{bmatrix}-c_1\!&\!-c_2\end{bmatrix}\!\bm{X}\!\leq\!-\sin(\psi_l\!-\!\psi_b)M\!-\!\alpha_1\epsilon_{\psi} \\
&~~~~~~~~~~~~~~~~\!~~~~+\begin{bmatrix}c_1&c_2\end{bmatrix}\bm{\tau}^{r\!e\!f}.\end{array} \label{eq18}
\end{equation}
where $\bm{X}=\bm{\tau}-\bm{\tau}^{r\!e\!f}$.

\subsubsection{CBF for CC-2}
For this constraint, we propose the following barrier function:
\begin{equation}
h(\bm{\nu}_l)=u-\epsilon_u, \label{eq19}
\end{equation}
where $\epsilon_u>0$ denotes another constant design parameter specifying the safety margin from singularities.

It is evident that this barrier function has relative degree 1, and the corresponding CBF can be selected as following \cite{b18}--\hspace{1sp}\cite{b20}
\begin{equation}
\dot{h}(\bm{\nu}_l,\dot{\bm{\nu}}_l)\geq-\alpha(h(\bm{\nu}_l)), \label{eq20}
\end{equation}
where $\alpha(\cdot)$ denotes any class-$\mathcal{K}$ function.

\emph{Proposition 2}. CBF condition for CC-2 can be expressed as follows:
\begin{equation}
\tau_u\geq-\dfrac{f_u+\alpha(u-\epsilon_u)}{b_u}. \label{eq21}
\end{equation}

\emph{Proof}. For a given $\alpha(u-\epsilon_u)$, it is always feasible to design $\tau_u$ such that the condition (\ref{eq21}) is satisfied, which in turn guarantees (\ref{eq20}) and ensures that the barrier function $h(\bm{\nu}_l)$ in (\ref{eq19}) is a CBF. This completes the proof. \qed

(\ref{eq21}) can be modified as following to match the constraint format required by \emph{quadprog($\cdot$)} in MATLAB,
\begin{equation}
\begin{bmatrix}-1&0\end{bmatrix}\bm{X}\leq \dfrac{f_u+\alpha(u-\epsilon_u)}{b_u}+\tau_u^{r\!e\!f}. \label{eq22}
\end{equation}

By combining (\ref{eq18}) and (\ref{eq22}), we finally obtain the following form of the constraints
\begin{equation}
\begin{array}{ll}
\text{if} & -\pi/2<\psi_l-\psi_b<\pi/2\\
&\bm{A}_1\bm{X}\leq \bm{b}_1, \\
\text{elseif} & \psi_l-\psi_b>\pi/2~\text{or}~\psi_l-\psi_b<-\pi/2\\
&\bm{A}_2\bm{X}\leq \bm{b}_2,\end{array} \label{eq23}
\end{equation}
where
\begin{align*}
\bm{A}_1&=\begin{bmatrix}c_1&c_2\\-1&0\end{bmatrix},~\bm{A}_2=\begin{bmatrix}-c_1&-c_2\\-1&0\end{bmatrix} \\
\bm{b}_1&=\begin{bmatrix}\sin(\psi_l-\psi_b)M-\alpha_1\epsilon_{\psi}-\begin{bmatrix}c_1&c_2\end{bmatrix}\bm{\tau}^{r\!e\!f}\\ [f_u+\alpha(u-\epsilon_u)]/b_u+\tau_u^{r\!e\!f}\end{bmatrix} \\
\bm{b}_2&=\begin{bmatrix}-\sin(\psi_l-\psi_b)M-\alpha_1\epsilon_{\psi}+\begin{bmatrix}c_1&c_2\end{bmatrix}\bm{\tau}^{r\!e\!f}\\ [f_u+\alpha(u-\epsilon_u)]/b_u+\tau_u^{r\!e\!f}\end{bmatrix}
\end{align*}

\subsection{Design of Reference Controller $\bm{\tau}^{r\!e\!f}$}
We now recall the two-input-two-output tracking model given in (\ref{eq9}) and (\ref{eq10}), and design a controller, which will serve as the reference controller $\bm{\tau}^{ref}$, using general backstepping method \cite{b13}. In the control design, it is assumed that both CC-1 and CC-2 are satisfied.

\subsubsection{Kinematic Tracking}
As mentioned before, to avoid the possible singularity of SP-4, this paper adopts the towing-type tracking method presented in \cite{b14} and force $p_e(t)\rightarrow c_d$ with $c_d>0$ a design parameter. With this in mind, this step considers the following Lyapunov function candidate
\begin{equation}
V_1=\dfrac{1}{2}\left[(p_e-c_d)^2+\gamma_{\psi}\psi_{le}^2\right], \label{eq24}
\end{equation}
where $\gamma_{\psi}>0$ is a weighting factor.

Differentiating (\ref{eq24}) and substituting (\ref{eq9}) into it, we obtain
\begin{align}
\dot{V}_1=&(p_e-c_d)\left[u_{ld}\cos(\psi_{ld}-\psi_b)-u_l\cos(\psi_l-\psi_b)\right] \nonumber \\
&+\gamma_{\psi}\psi_{le}(\dot{\psi}_{ld}-r_l). \label{eq25}
\end{align}

According to (\ref{eq25}), the control laws for the stabilizing functions of the virtual inputs $u_l$ and $r_l$ are chosen as follows:
\begin{align}
\alpha_{u_l}&=\sec(\psi_l\!-\!\psi_b)\left[u_{ld}\cos(\psi_{ld}\!-\!\psi_b)\!+\!k_p(p_e\!-\!c_d)\right], \label{eq26} \\
\alpha_{r_l}&=\dot{\psi}_{ld}+\gamma_{\psi}^{-1}k_{\psi}\psi_{le}, \label{eq27}
\end{align}
where $k_p,k_{\psi}>0$ are design parameters.

By substituting (\ref{eq26}) and (\ref{eq27}) into (\ref{eq25}), we have
\begin{align}
\dot{V}_1=&-k_p(p_e-c_d)^2-k_{\psi}\psi_{le}^2+(p_e-c_d)e_{u_l}\cos(\psi_l-\psi_b) \nonumber \\
&+\gamma_{\psi}\psi_{le}e_{r_l}, \label{eq28}
\end{align}
where $e_{u_l}=\alpha_{u_l}-u_l$ and $e_{r_l}=\alpha_{r_l}-r_l$.

\subsubsection{Dynamic Tracking}
The vessel's dynamics (\ref{eq10}) can be rewritten in the following error form
\begin{equation}
\dot{\bm{e}}=\begin{bmatrix}\dot{\alpha}_{u_l}-f_{u_l} \\ \dot{\alpha}_{r_l}-f_{r_l}\end{bmatrix}-\begin{bmatrix}b_{u_l} &\epsilon_{ra} \\0 &b_r\end{bmatrix}\bm{\tau}, \label{eq29}
\end{equation}
where $\bm{e}=[e_{u_l},e_{r_l}]^T$. Correspondingly, the Lyapunov function candidate in this step is chosen as follows:
\begin{equation}
V_2=V_1+\dfrac{1}{2}\bm{e}^T\bm{G}\bm{e}, \label{eq30}
\end{equation}
where $\bm{G}=\text{diag}(\gamma_u,\gamma_r)$ and $\gamma_u,\gamma_r>0$ are weighting factors.

Differentiating (\ref{eq30}) yields
\begin{align}
\dot{V}_2=&-k_p(p_e-c_d)^2\!-\!k_{\psi}\psi_{le}^2\!+\!(p_e\!-\!c_d)e_{u_l}\cos(\psi_l\!-\!\psi_b) \nonumber \\
&+\gamma_{\psi}\psi_{le}e_{r_l}\!+\!\bm{e}^T\bm{G}\!\left(\begin{bmatrix}\dot{\alpha}_{u_l}\!-\!f_{u_l}\\ \dot{\alpha}_{r_l}\!-\!f_{r_l}\end{bmatrix}\!\!-\!\!\begin{bmatrix}b_{u_l}\!&\!\epsilon_{ra}\\0\!&\!b_r\end{bmatrix}\!\bm{\tau}\right)\!\!. \label{eq31}
\end{align}

Consequently, according to (\ref{eq31}), the reference controller can be designed as follows:
\begin{align}
\bm{\tau}^{r\!e\!f}=&\begin{bmatrix}b_{u_l}&\epsilon_{ra}\\0&b_r\end{bmatrix}^{-1}\left[\begin{bmatrix}\dot{\alpha}_{u_l}-f_{u_l}\\ \dot{\alpha}_{r_l}-f_{r_l}\end{bmatrix} +\bm{G}^{-1}\left(\bm{K}\bm{e}\begin{matrix} \\ \\ \end{matrix} \right.\right. \nonumber \\
&\left.\left.+\begin{bmatrix}(p_e-c_d)\cos(\psi_l-\psi_b)\\ \gamma_{\psi}\psi_{le}\end{bmatrix}\right)\right], \label{eq32}
\end{align}
where $\bm{K}=\text{diag}(k_u,k_r)$ with $k_u,k_r>0$ design parameters.

It is straightforward to see that whether $\epsilon_{ra}$ is zero or not (i.e., whether the system is minimum-phase or non-minimum phase) does not affect the computation of the reference control law $\bm{\tau}^{r\!e\!f}$.

Substituting (\ref{eq32}) with $\bm{\tau}=\bm{\tau}^{ref}$ into (\ref{eq31}) yields
\begin{align}
\dot{V}_2&=-k_p(p_e-c_d)^2-k_{\psi}\psi_{le}^2-k_ue_{u_l}^2-k_re_{r_l}^2 \nonumber \\
&\leq -\lambda V_2, \label{eq33}
\end{align}
where $\lambda:=0.5\cdot min\{1/k_p,\gamma_{\psi}/k_{\psi},\gamma_u/k_u,\gamma_r/k_r\}$.

\emph{Theorem 1}. Consider the trajectory tracking problem governed by (\ref{eq9}) and (\ref{eq10}) with CC-1 and CC-2 satisfied. If the control input is designed as $\bm{\tau}^{ref}$ in (\ref{eq30}), then the closed-loop tracking system is guaranteed to be exponentially stable in terms of polar coordinates.

\emph{Proposition 3}. By appropriately selecting $c_d,\gamma_{\psi},\gamma_u,\gamma_r>0$, it is always possible to guarantee $p_e(t)>0$ for all $t\geq 0$, given any initial conditions $p_e(0)>0$, $\psi_{le}(0)$, $e_{u_l}(0)$, and $e_{r_l}(0)$.

\emph{Proof}. For the given initial conditions, let $c_d$ be designed such that
\begin{equation}
c_d\geq\dfrac{0.5p_e^2(0)+V_R(0)}{p_e(0)}, \label{eq34}
\end{equation}
where $V_R(t)=\gamma_{\psi}\psi_{le}^2(t)+\gamma_ue_{u_l}^2(t)+\gamma_r e_{r_l}^2(t)$. Further by expanding (\ref{eq34}), we obtain
\begin{equation}
\dfrac{1}{2}c_d^2\geq\dfrac{1}{2}\left[p_e(0)-c_d\right]^2+V_R(0)=V_2(0). \label{eq35}
\end{equation}

Now, suppose that $p_e(t)=0$ at some $t=t_c\geq0$. In this case, we have
\begin{equation}
V_2(t_c)=\dfrac{1}{2}c_d^2+V_R(t_c)\geq \dfrac{1}{2}c_d^2. \label{eq36}
\end{equation}

Combining (\ref{eq36}) with (\ref{eq35}), it follows that $V_2(t_c)\geq V_2(0)$, which contradicts (\ref{eq33}). Therefore, to avoid this contradiction, the only possible case is $p_e(t)>0$ for all $t\geq0$, which concludes the proof. \qed

\emph{Remark 6}. By choosing sufficiently small values of $\gamma_{\psi}$, $\gamma_u$, and $\gamma_r$, the condition in (\ref{eq34}) can be approximated as $c_d\geq 0.5p_e(0)$. It is noteworthy, however, that (\ref{eq34}) represents a sufficient condition for ensuring $p_e(t)>0$ for all $t\geq 0$, rather than a necessary one.

\subsection{QP-Based Tracking Controller Design}
Consequently, the proposed QP-based trajectory tracking controller for underactuated marine vessels, whose kinematics and dynamics are governed by (\ref{eq1}) and (\ref{eq2}), is formulated simply as follows:
\begin{align}
\underset{\dot{\bm{\nu}}_l,\bm{\tau}}{\text{min}}~~&||\bm{X}||^2 \label{eq37} \\
\text{s.t.}~~&\text{Constraint (\ref{eq23})}, \nonumber
\end{align}
where $\bm{X}=\bm{\tau}-\bm{\tau}^{r\!e\!f}$ with $\bm{\tau}^{r\!e\!f}$ defined as (\ref{eq32}).

\emph{Assumption 1} (Assumption 2 in \cite{b20}). The control law $\bm{\tau}$ generated by QP (\ref{eq37}) is locally Lipschitz.

\emph{Remark 7}. This standard assumption is necessary to ensure that the barrier functions $h(\bm{\eta}_l)$ defined in (\ref{eq12}) and (\ref{eq15}) are ECBFs \cite{b17}--\hspace{1sp}\cite{b19}.

\section{Numerical Study}
This section presents numerical simulations to verify the effectiveness of the proposed control method. In this MATLAB simulation, we consider the trajectory tracking problem of an underactuated surface vessel, whose dynamics are described as follows \cite{b5}:
\begin{align}
\dot{u}\!\!&\!=\!-\!\dfrac{d_u}{m_{11}}\!+\!\dfrac{m_{22}}{m_{11}}vr\!-\!\dfrac{d_{u2}}{m_{11}}|u|u\!-\!\dfrac{d_{u3}}{m_{11}}u^3\!+\!\dfrac{1}{m_{11}}\tau_u,\nonumber \\
\dot{v}\!\!&=\!-\!\dfrac{d_v}{m_{22}}v\!-\!\dfrac{m_{11}}{m_{22}}ur\!-\!\dfrac{d_{v2}}{m_{22}}|v|v\!-\!\dfrac{d_{v3}}{m_{22}}v^3,\label{eq38} \\
\dot{r}\!\!&=\!\!-\!\!\dfrac{d_r}{m_{33}}r\!\!+\!\!\dfrac{(m_{11}\!\!-\!\!m_{22})}{m_{33}}ur\!\!-\!\!\dfrac{d_{r2}}{m_{33}}|r|r\!\!-\!\!\dfrac{d_{r3}}{m_{33}}r^3\!\!+\!\!\dfrac{1}{m_{33}}\tau_r,~~~~~~ \nonumber
\end{align}
where $m_{11}=1.2e+5$, $m_{22}=1.779e+5$, $m_{33}=6.36e+7$, $d_u\!=\!2.152+4$, $d_v\!=\!1.47+5$, $d_r\!=\!8.02e+6$, $d_{u2}\!=\!0.2d_u$, $d_{u3}\!\!=\!\!0.1d_u$, $d_{v2}\!\!=\!\!0.2d_v$, $d_{v3}\!\!=\!\!0.1d_v$, $d_{r2}\!\!=\!\!0.2d_r$, $d_{r3}\!\!=\!\!0.1d_r$.

In the trajectory tracking, the reference trajectory is taken as: if $t\in[0,60s)$, $u_{ld}(t)=5m/s$ with $\bm{\eta}_d(0)=[100m,30m,0]^T$; otherwise, $u_{ld}(t)=5m/s$ and $\dot{\psi}_{ld}(t)=-0.05rad/s$. The vessel's initial conditions are $\bm{\eta}(0)=[90m,25m,30deg]^T$ and $\bm{\nu}(0)=[1,0,0]^T$, and the controller design parameters are set as: $k_p=1$. $k_{\psi}=6$, $k_u=3$, $k_r=1$, $\gamma_{psi}=\gamma_u=\gamma_r=1$, $c_d=6m$. CBF related parameters are chosen as $\epsilon_{\psi}=15deg$ and $\epsilon_u=0.5m/s$, and for $\bm{K}_{\alpha}$ in (\ref{eq13}), we choose $\alpha_1=0.01$ and $\alpha_2=0.3$, which ensure that $\bm{F}-\bm{G}\bm{K}_{\alpha}$ is Hurwitz.

Also, as in \cite{b6,b7}, the following low-pass filter is employed to calculate $\ddot{\bm{\nu}}=[\ddot{u},\ddot{v},\ddot{r}]^T$, which is used to compute $\ddot{\psi}_a(t)$,
\begin{equation}
\ddot{\bm{\nu}}(t_{k+1})=(1\!-\!\mu)\ddot{\bm{\nu}}(t_k)\!+\!\mu \ddot{\bm{\nu}}^m(t_{k+1}),~k=0,1,\cdots, \label{eq39}
\end{equation}
where the filter order is set as $\mu=2^{-3}$ and $\ddot{\bm{\nu}}(t_{k+1})=(\dot{\bm{\nu}}(t_{k+1})-\dot{\bm{\nu}}(t_k))/T_s$ with $T_s$ the sampling time and taken as $T_s=0.01s$ in the simulation. Here $\dot{\bm{\nu}}$ is calculated by the vessel's dynamics (\ref{eq38}).

In the simulation, two cases are compared: (i) trajectory tracking performed by directly applying $\bm{\tau}=\bm{\tau}^{r\!e\!f}$, where $\bm{\tau}^{r\!e\!f}$ is computed as in (\ref{eq32}); and (ii) trajectory tracking using the proposed method described in (\ref{eq37}).

\begin{figure}[!t]
\centerline{\includegraphics[width=\columnwidth]{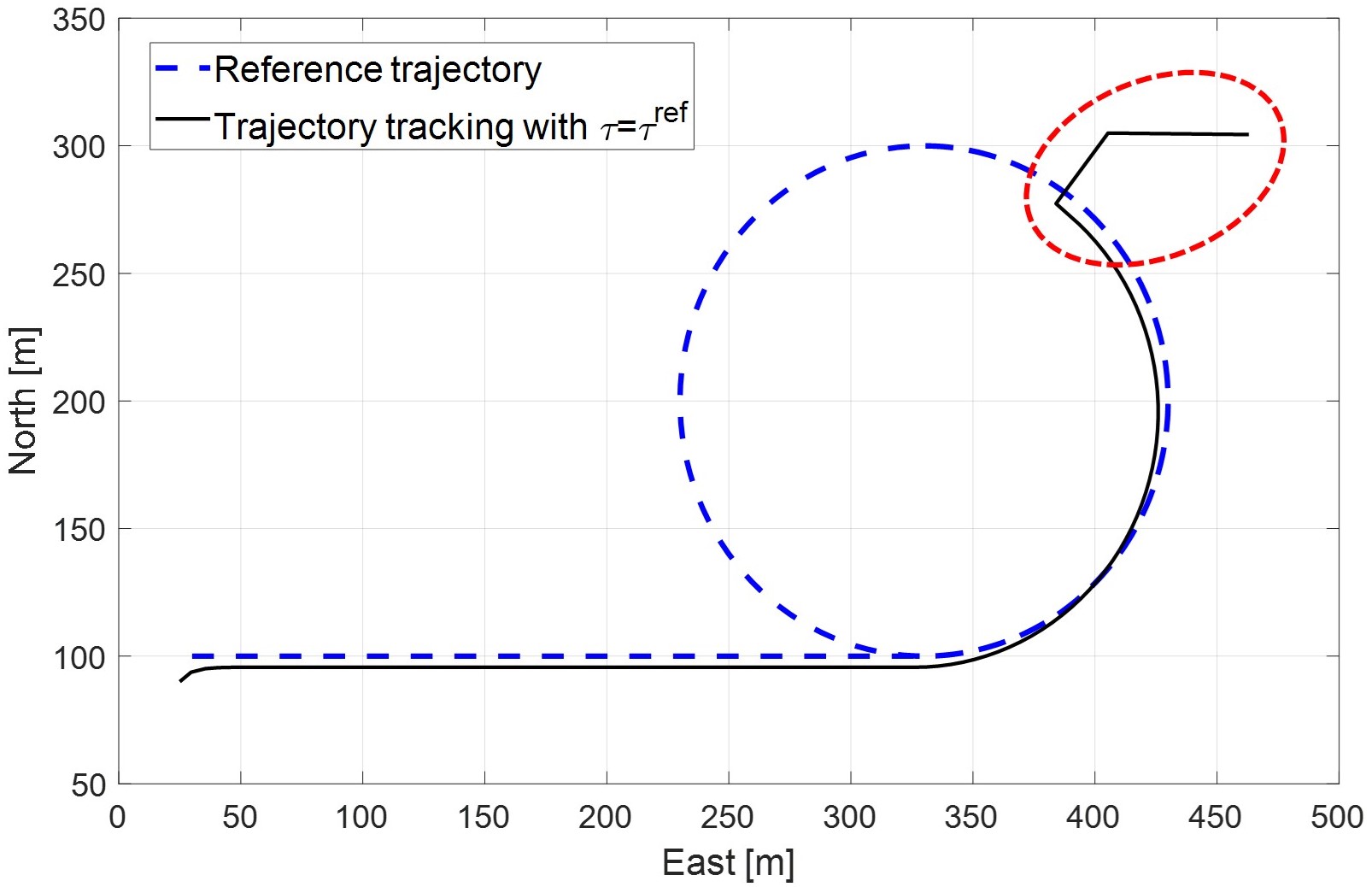}}
\caption{Reference trajectory and its tracking using $\bm{\tau}=\bm{\tau}^{r\!e\!f}$.}
\label{fig1}
\end{figure}

\begin{figure}[!t]
\centerline{\includegraphics[width=\columnwidth]{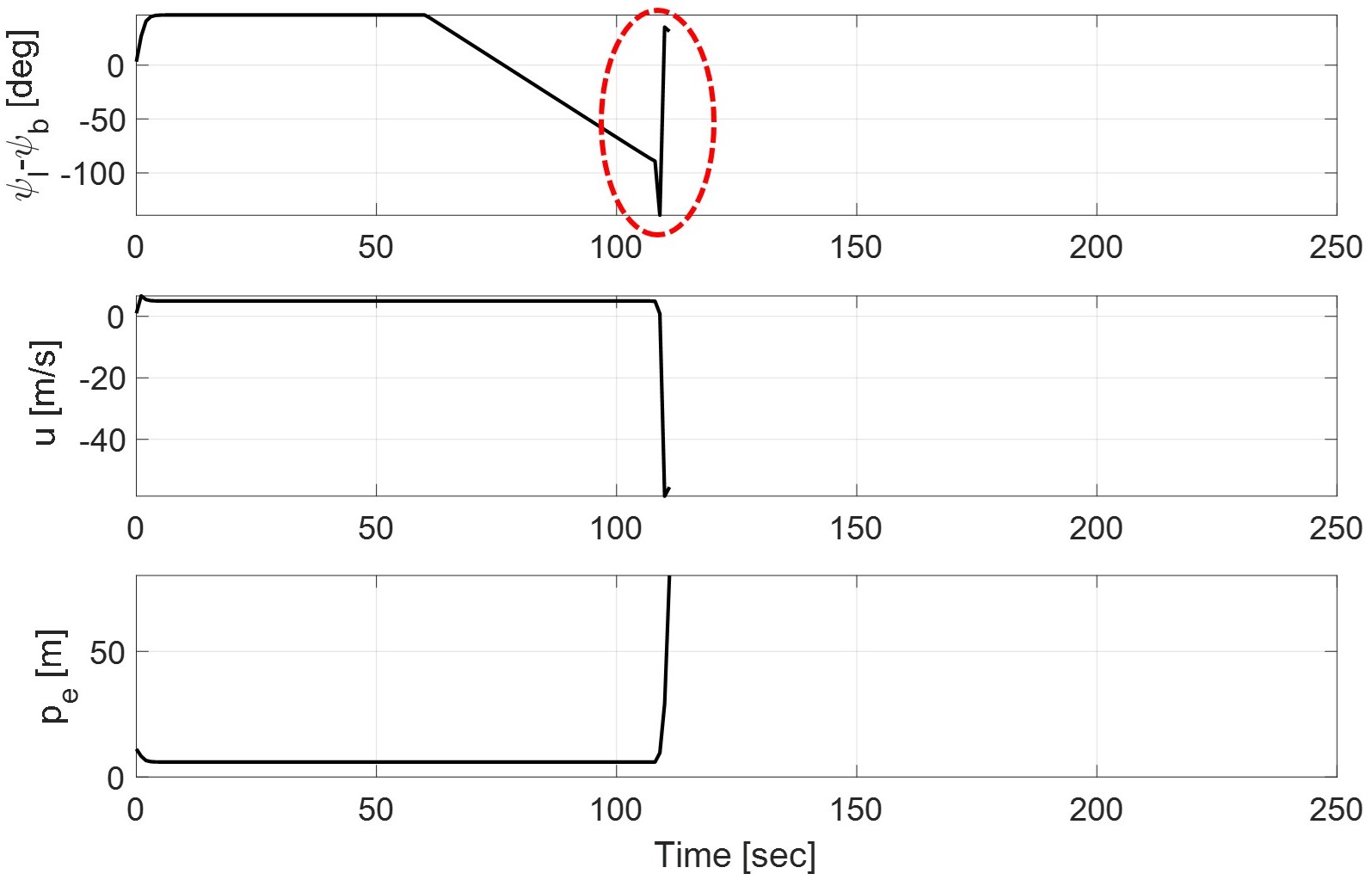}}
\caption{Variation trends of all possible causes associated with SP-1 to SP-4.}
\label{fig2}
\end{figure}

The simulation results for case (i) are shown in Fig. \ref{fig1} and \ref{fig2}, from which it can be observed that trajectory tracking fails to be properly achieved. In fact, the control system breaks down at approximately $t\approx 110s$. From Fig. \ref{fig2}, it can be seen that for $t<110s$, $u\approx 5m/s$ and $p_e\approx 6$. Therefore, SP-2 to SP-4 do not correspond to problematic situations, and the only possible case is that $\psi_l-\psi_b$ approaches $-\pi/2$, leading to $\cos(\psi_l-\psi_b)\approx 0$. Consequently, the only singularity encountered in case (i) is SP-1.

\begin{figure}[!t]
\centerline{\includegraphics[width=\columnwidth]{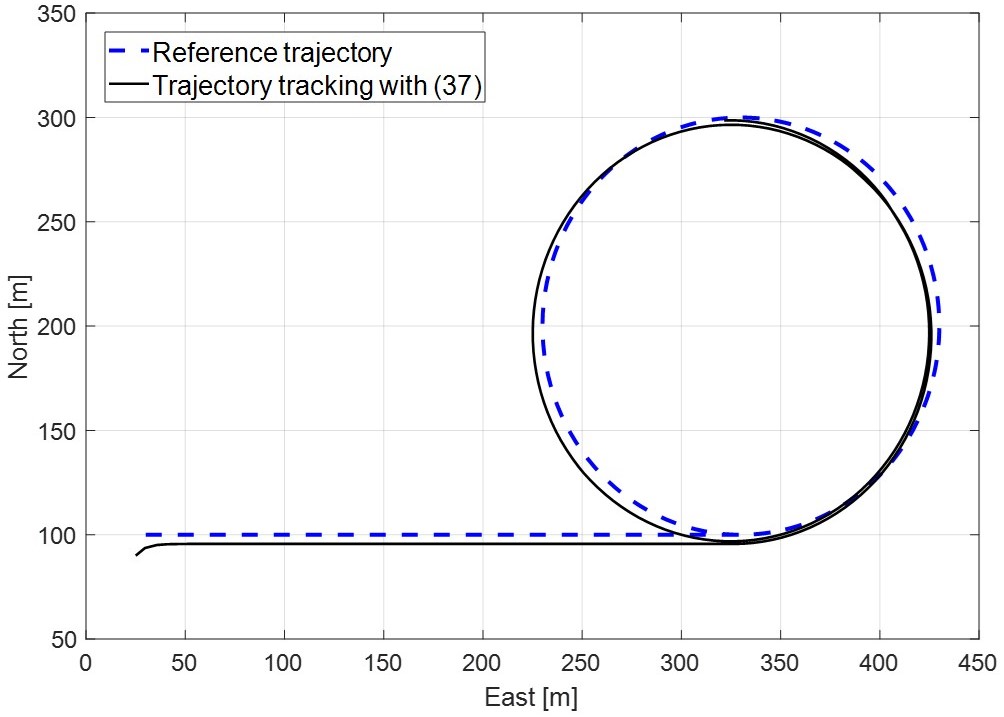}}
\caption{Reference trajectory and its tracking using the proposed method in (\ref{eq37}).}
\label{fig3}
\end{figure}

\begin{figure}[!t]
\centerline{\includegraphics[width=\columnwidth]{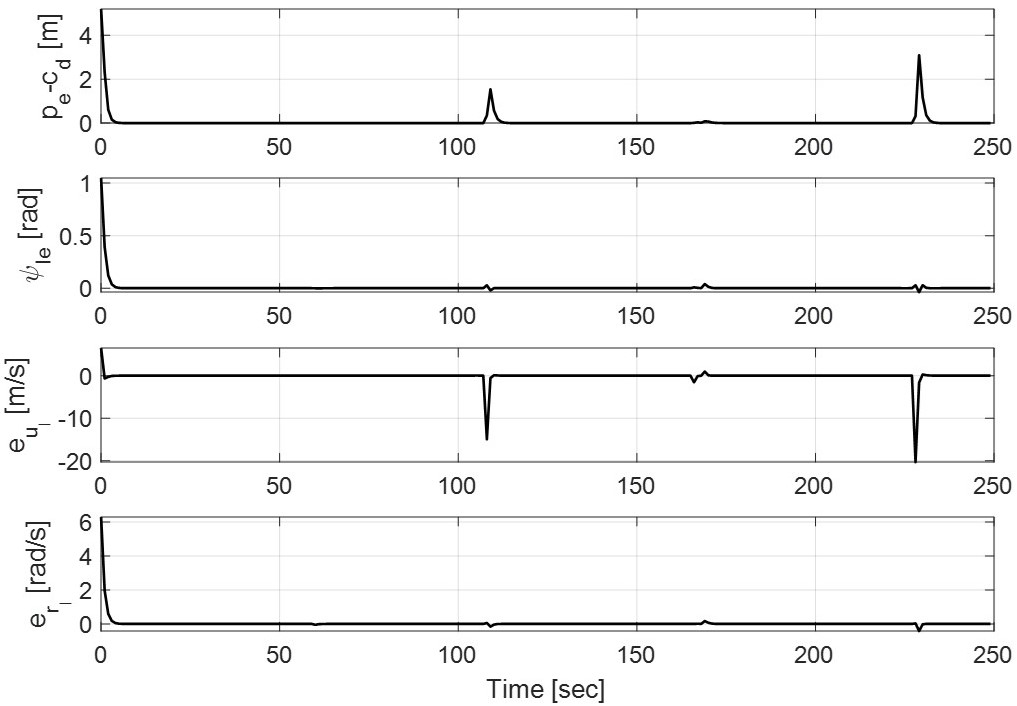}}
\caption{Tracking error convergence with the proposed method (\ref{eq37}).}
\label{fig4}
\end{figure}

\begin{figure}[!t]
\centerline{\includegraphics[width=\columnwidth]{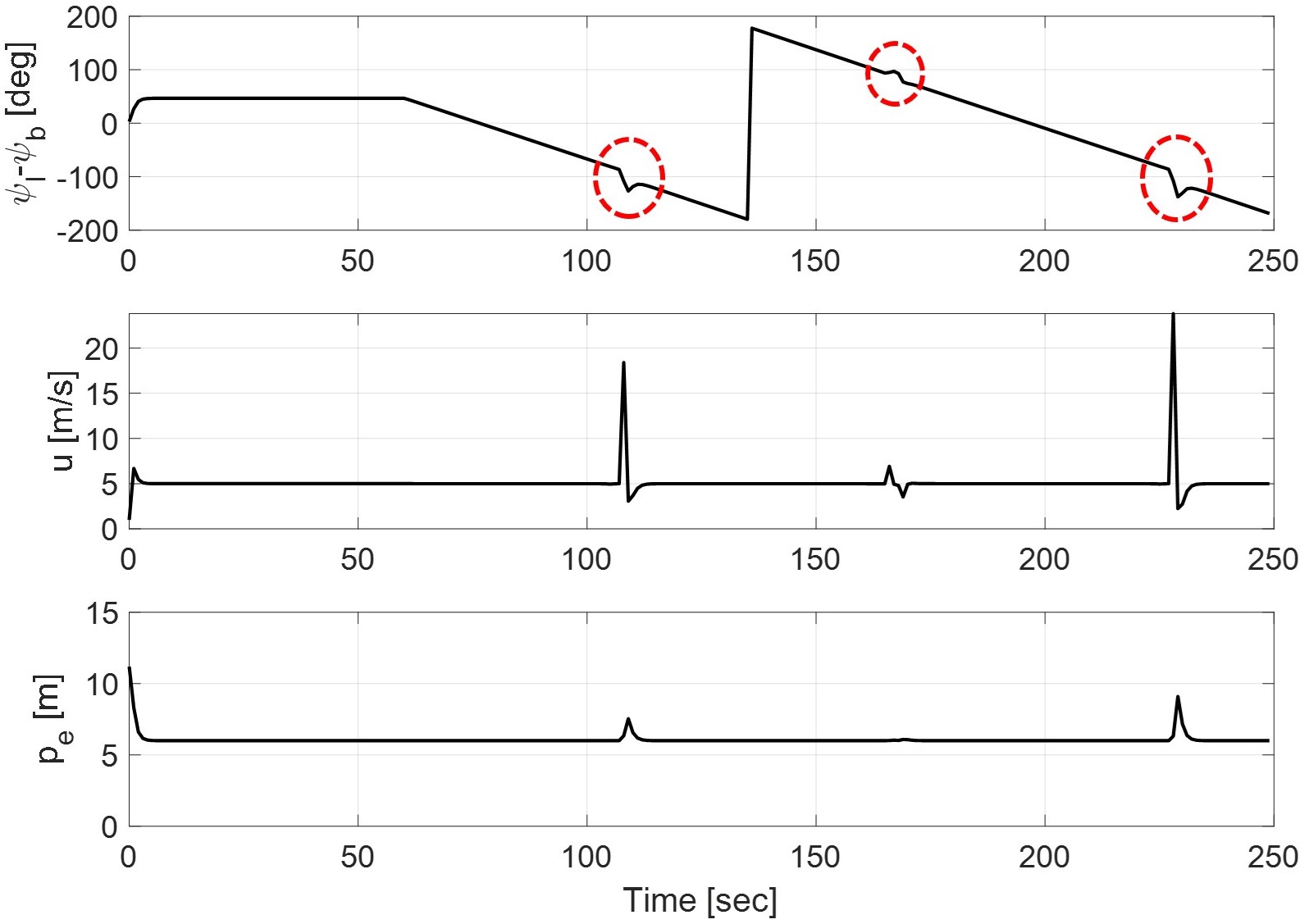}}
\caption{Variation trends of all possible causes associated with SP-1 to SP-4 in the case of the proposed method (\ref{eq37}).}
\label{fig5}
\end{figure}

In contrast, as shown in Figs. \ref{fig3} and \ref{fig4}, case (ii), in which the proposed method is applied, successfully tracks the given reference trajectory. For a more in-depth analysis, all three potential causes related to SP-1 through SP-4 are plotted in Fig. \ref{fig5}. Similar to case (i), during the entire tracking process, we have $u\approx 5m/s$ and $p_e\approx 6$. The issue is that $\psi_l-\psi_b$ still crosses the potential singular points, which are indicated by the ref dashed circles in Fig. \ref{fig5}. Fortunately, the proposed QP-based control method functions properly at these points, ensuring continuous trajectory tracking. In particular, it can be observed that the proposed method causes $\psi_l-\psi_b$ to make jump near these points to avoid the singularities. This characteristic may well represent the most significant advantage of employing the QP-based approach in this paper.

It is worth noting that, since the reference trajectory follows a circular motion in the simulation, it is seemingly unavoidable for $psi_l-\psi_b$ to across these singulary points in order for $\psi_{le}=\psi_{ld}-\psi_l\rightarrow 0$. Fortunately, the proposed QP optimizer enables continuous tracking of the vessel by allowing the system to effectively jump over or bypass these singularity points.

\begin{figure}[!t]
\centerline{\includegraphics[width=\columnwidth]{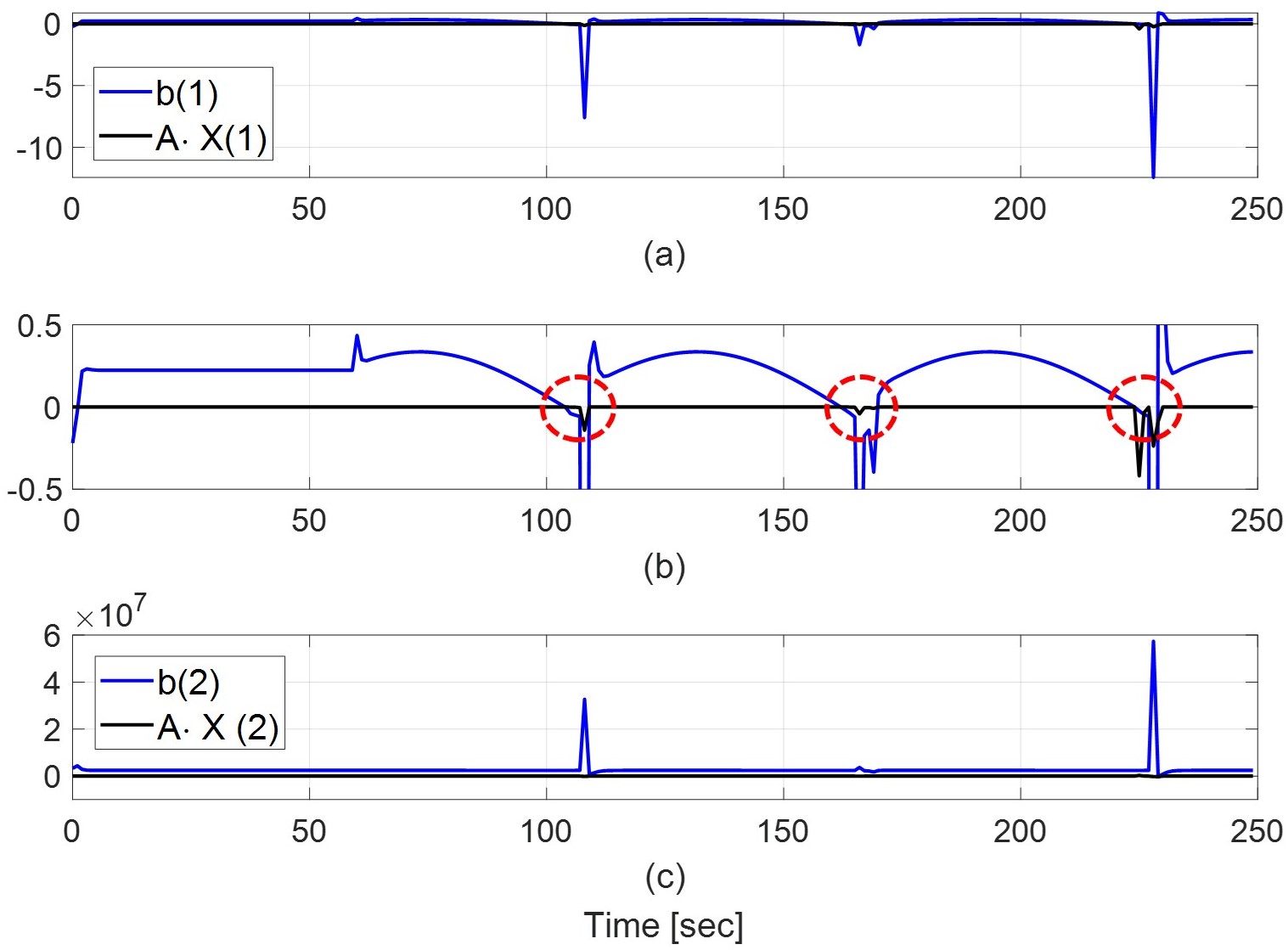}}
\caption{Verification of constraint (\ref{eq23}) in the case of the proposed method (\ref{eq37}).}
\label{fig6}
\end{figure}

\begin{figure}[!t]
\centerline{\includegraphics[width=\columnwidth]{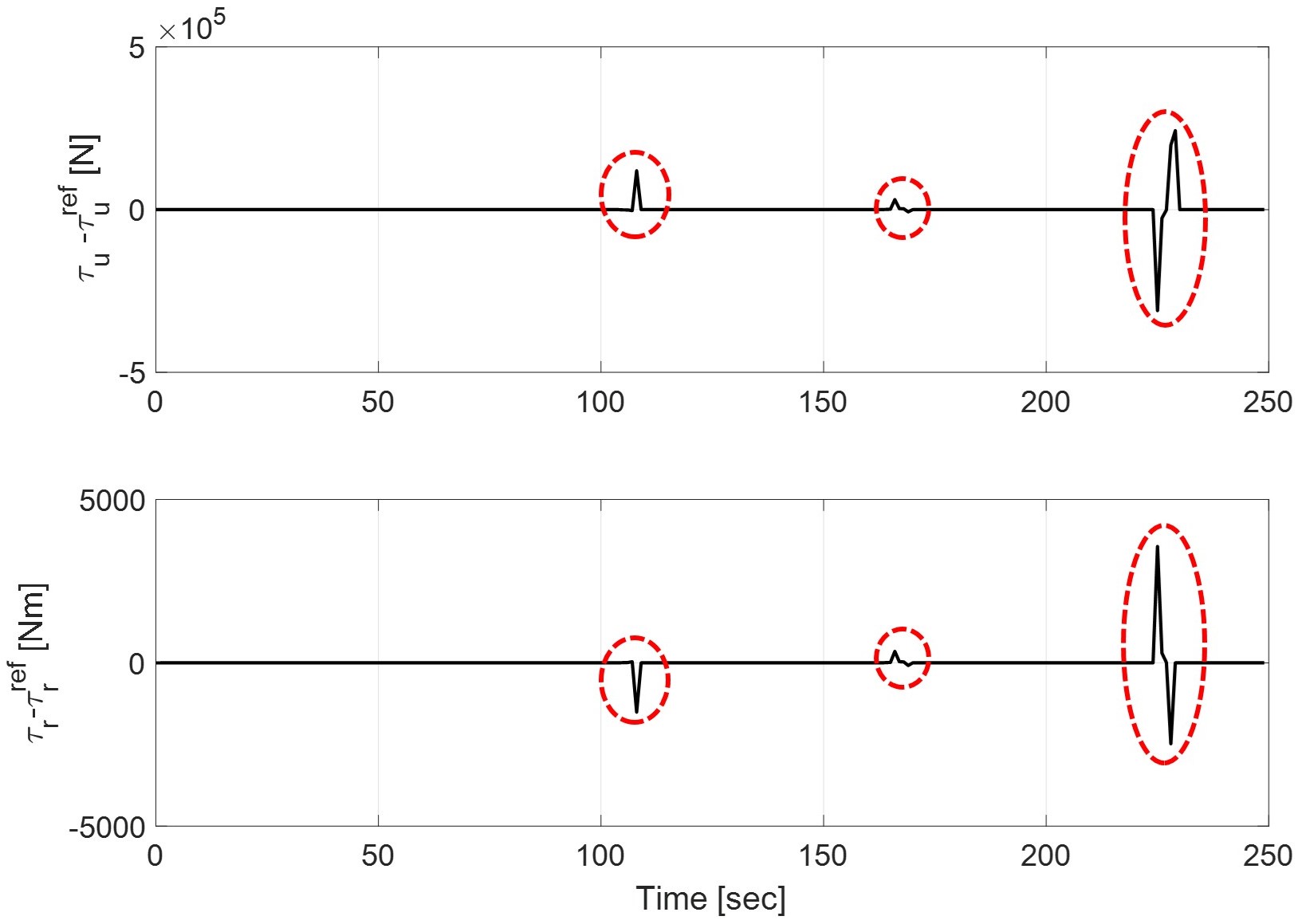}}
\caption{Variation trends of $\bm{X}=\bm{\tau}-\bm{\tau}^{r\!e\!f}$.}
\label{fig7}
\end{figure}

Fig. \ref{fig6} illustrates whether the constraint conditions expressed in (\ref{eq23}) are satisfied during the tracking process. As can be seen, when $\psi_l-\psi_b$ is near the singular points, CC-1 is not satisfied. Correspondingly, the variation of $\bm{X}=\bm{\tau}-\bm{\tau}^{r\!e\!f}$ is shown in Fig. \ref{fig7}. As observed, $\bm{X}$ takes nonzero values only when $\psi_l-\psi_b$ is near the singular points, while it remains zero otherwise.

From the above simulation results, it can be confirmed that the proposed QP-based control scheme effectively overcomes the singularities considered in this paper, particularly the one caused by $\cos(\psi_l-\psi_b)=0$.

\section{Conclusion}
This paper has presented a CBF based QP method for the trajectory tracking of underactuated marine vessels. Two polar coordinates transformations were employed to convert the original two-input-three-output vessel model into a two-input-two-output feedback form. To overcome the potential singularities caused by these transformations, the proposed method adopts CBF framework instead of the AMO and EMO concepts used in existing studies. Moreover, the restrictive assumption in previous works -- that the vessel's surge velocity must always remain strictly positive -- is no longer required here, as it is incorporated as a constraint within the QP solver.

Numerical simulations have shown that the control inputs tend to increase significantly near singular points to avoid singularities. However, in practical implementations, such excessively large inputs may be infeasible due to actuator limitations (e.g., finite thrust force and yaw moment). Therefore, an interesting direction for future work is to investigate how the proposed method can be extended to effectively handle control input saturation.




\end{document}